\documentstyle[revtex]{aps}
\widetext
\begin{document}
\draft
\begin{title}
Search for SUSY with R-Parity Violation at $\gamma p$ and $\gamma e$ Colliders
\end{title}
\author{Z.Z.Aydin, A. Kandemir and A.U. Y\i lmazer}
\begin{instit}
Ankara University, Faculty of Sciences \\
Department of Engineering Physics \\
06100 Tando\u{g}an, Ankara - Turkey
\end{instit}
\begin{abstract}
We present an outlook for possible $R_p$ violating SUSY search at
$\gamma p$ and $\gamma e$ colliders. Single production of third generation
slepton/sneutrino through the $\lambda'_{ijk}$ couplings at a $\gamma p$
collider is investigated and compared
with the results of resonant sfermion productions at the existing colliders.
Also single sneutrino production at a future $\gamma e$ collider through the
$\lambda _{[ij]k}$ couplings is discussed.
\end{abstract}

\section{Introduction}
\label{sec:intro}

   Among the various extensions beyond the Standard Model (SM), supersymmetry
(SUSY) appears to be a well-motivated strong option to investigate the physics
at TeV scale. In the past most discussions of SUSY phenomenology assumed
R-parity ($R_p$) conservation \cite{haber86}. R-parity is defined by

\begin{eqnarray}
R_p=(-1)^{3B+L+2S}
\end{eqnarray}
where $S$ is the spin, $B$ is the baryon-number and $L$ is the lepton-number
of the particle. Hence under this discrete symmetry SM particles are even
while their superpartners are odd. This implies that SUSY particles
are pair produced, every SUSY particle decays into another one and that
there is lightest supersymmetric particle (LSP) which is stable.
Actually conservation of R-parity is put by purpose to ensure that the minimal
supersymmetric extension of the standard model (MSSM) retain the symmetries
of the SM, where in particular baryon and lepton quantum numbers are conserved
separately \cite{fayet78}.

In the usual formulation the MSSM is defined by the superpotential

\begin{eqnarray}
W_{R} = Y^u_{ij} Q_i \cdot H_u U_j + Y^d_{ij} Q_i \cdot H_d \bar
D_j + Y^e_{ij} L_i \cdot H_d \bar E_j + \mu H_u \cdot H_d.
\end{eqnarray}
which respects the multiplicative R-parity defined above. However there is no
deep theoretical justification for imposing $R_p$ conservation, since gauge and Lorentz
invariances allow for the following additional terms in the superpotential

\begin{eqnarray}
W_{{R\!\!\!/}_p} = {1\over2} \lambda_{[ij]k} L_i \cdot L_j \bar E_k +
\lambda'_{ijk} L_i \cdot Q_j \bar D_k + {1\over2}
\lambda^{\prime\prime}_{i[jk]} \bar U_i \bar D_j \bar D_k + \epsilon_i
L_i \cdot H_u,
\end{eqnarray}
which explicitly break it. The $MSSM$ with $R_p$ conservation has been
extensively studied and direct searches for superpartners at the existing
colliders are still continuing \cite{hep99}. On the other hand $R_p$
violation leads
to a different phenomenology and recently the interest in the
${R\!\!\!/}_p$ models has been motivated by the observation of a number
of events at high $Q^2$ at  HERA. Also in $R_p$-violating models neutrinos
acquire masses and can mix; therefore it is a good candidate
to explain the neutrino oscillations observed in
Super-Kamiokande \cite{fukuda98}.
Terms with the lepton number violating
Yukawa couplings $\lambda'_{ijk}$ make the ep collider HERA especially
promising for the resonant production of squarks,
and possible explanations of these high $Q^2$ events within
the framework of R-parity violating supersymmetry have been
discussed in the literature \cite{chou97}. Although further data
taken by the H1 and ZEUS collaborations failed to produce any further
"excess" events with high-$Q^2$ a rich phenomenology of ${R\!\!\!/}_p$
emerged for HERA. Also formation of s-channel slepton and squark
resonances at LEP2 and TEVATRON at current energies is an exciting
possibility in ${R\!\!\!/}_p$-$SUSY$ searches \cite{barbier98}.

On the other hand in recent years in addition to
the existing colliders the possibilities
of the realization of $\gamma e$, $\gamma \gamma$ and $\gamma p$
colliders have been proposed and discussed in detail \cite{aydin96}.
Colliding the
beam of high energy photons produced by Compton backscattering of
laser photons off linac electrons with
the beam of a proton ring is the idea leading
to TeV scale $\gamma p$ colliders. Physics program of the photon
colliders are studied in \cite{telnov00}.
Also search for SUSY in polarized  $\gamma p$
collisions have been investigated in \cite{yilmazer96}.

In searching the superpartners
squarks might be too heavy to be produced at HERA, LEP or
TEVATRON, but sleptons are generally expected to be lighter than squarks
so single slepton production would be interesting. Also pair production of
sleptons via $R_p$ conserving mechanisms might be closed kinematically.
The s-channel slepton resonance production
via ${R\!\!\!/}_p$ interactions in $e^+ e^-$
collisions through
$e^+e^-\rightarrow \tilde{\nu}\rightarrow l^+l^-$, and
in $p\bar p$ collisions through
$p\bar p \rightarrow \tilde{\nu}\rightarrow l^+l^-$,
$p\bar p \rightarrow \tilde{l}^+ \rightarrow l^+ \nu$ have been examined
in the literature  \cite{lep99}. In $e^+p$ collisions at HERA squark resonance
productions via
$e^+d^k_R \rightarrow \tilde{u}^j_L$  ($\tilde{u}^j=
\tilde{u}, \tilde{c}, \tilde{t}$),
$e^+\bar{u}^j_L \rightarrow \bar {\tilde{d^k}}_R$,
($\tilde{d}^k=\tilde{d}, \tilde{s}, \tilde{b}$) have been investigated
in \cite{hera99}.

Although HERA, LEP, FERMILAB and LHC should be sufficient to check
the low energy SUSY however
experiments at all possible types of colliding beams would be inevitable
to explore the new physics around the TeV scale, and hence future
$\gamma e$ and $\gamma p$ colliders might play
a complementary role to the existing facilities.
In  this article we mainly focus on the single stau and sneutrino
productions at photon-proton and photon-electron collisions
as an alternative to the above-mentioned s-channel
resonance production.

\section{Selectron(stau) production at gamma-proton colliders}
\label{sec:pro}

In four-component Dirac notation the $R_p$ violating Lagrangian generated
by $W_{{R\!\!\!/}_p}$ is

\begin{eqnarray}
{\cal L}_{{R\!\!\!/}_p} &=& 
\lambda_{[ij]k} \Big[\tilde\nu_{iL} \bar
e_{kR} e_{jL} + \tilde e_{jL} \bar e_{kR} \nu_{iL} + \tilde
e^\star_{kR} \overline{(\nu_{iL})^C} e_{jL} - \tilde\nu_{jL} \bar
e_{kR} e_{iL} \nonumber \\[2mm] && - \tilde e_{iL}
\bar e_{kR} \nu_{jL} + \tilde e^\star_{kR} \overline{(\nu_{jL})^C}
e_{iL}\Big] + \lambda'_{ijk} \Big[\tilde\nu_{iL} \bar d_{kR}
d_{jL} + \tilde d_{jL} \bar d_{kR} \nu_{iL} \nonumber \\[2mm]
&& + \tilde d^\star_{kR} \overline{(\nu_{iL})^C} d_{jL} - \tilde e_{iL} \bar
d_{kR} u_{jL} - \tilde u_{jL} \bar d_{kR} e_{jL} - \tilde
d^\star_{kR} \overline{(e_{iL})^C} u_{jL}\Big] 
+ \lambda^{\prime\prime}_{i[jk]} \epsilon_{\alpha\beta\gamma} 
\nonumber \\[2mm] 
&& \Big[\tilde u^\star_{iR\alpha} \bar
d_{kR\beta} d^C_{jR\gamma} + \tilde d_{jR\beta} \bar
e_{kR\gamma} u^C_{iR\alpha} + \tilde d^\star_{kR\gamma}
\overline{(u_{iR\alpha})^C} d_{jR\beta}\Big] + h.c.
\end{eqnarray}
where $i, j, k$  are the generation indices. In photon-proton
collisions a single slepton or a single squark production
is possible. Let us take first
as an example, the single charged slepton production,
$\gamma p \rightarrow \tilde{e}^j_L X$.
One of the relevant subprocesses, $\gamma u \rightarrow \tilde{e}^j_L d^k_R$,
proceeds via the s-channel u-quark, t-channel slepton and u-channel d-quark
exchanges.
The invariant amplitude in four-component Dirac notation
(which could be written equally in two-component Weyl language) is

\begin{eqnarray}
M=N_c g_e \lambda' \epsilon_{\mu}(k)\bar{u}(p')Q^{\mu} u(p)\\[4mm]
Q^{\mu}={1\over2} (1-\gamma_5)\biggr [\frac{\not\! k+\not\! p}
{\hat s-m^2_u}\gamma ^{\mu}+
\frac {(p-p'+k')^{\mu}}{\hat t -{m_{\tilde e}}^2}
+\gamma ^{\mu}\frac{\not\! p'-\not\! k}
{\hat u-m^2_d}\biggr ]
\end{eqnarray}
where $g_e=\sqrt {4\pi \alpha}$, $\epsilon_{\mu}(k)$ is the
photon polarisation and $k, p, k'$ and $p'$ are the four momenta
of the photon, quark in the proton, slepton and outgoing quark
respectively. The differential cross-section for the subprocess is
given by
\begin{eqnarray}
\frac{d\hat{\sigma}}{d\hat{t}}=\frac{1}{16\pi\hat{s}^2}M^2
\end{eqnarray}
and after performing the integration over $\hat t$ one can easily obtain
the total cross-section for the subprocess
$\gamma u \rightarrow {\tilde e}^j_L d^k_R$. In order to obtain
the total cross-section for the process
$\gamma p \rightarrow {\tilde e}^i_L d^k_R X$ one should integrate
$\hat{\sigma}$ over the quark and photon distributions.
For this
purpose we make the following change of variables: first
expressing $\hat{s}$ as $\hat{s}=x_1x_2s$ where
$\hat{s}=s_{{\gamma}q}$, $s=s_{ep}$, $ x_1=E_{\gamma}/E_e$,
$x_2=E_q/E_p$ and furthermore calling $\tau =x_1x_2$, $x_2=x$
then one obtains $dx_1dx_2 = dx d\tau/x$. The limiting values
are $x_{1,max}=0.83$ in order to get rid of the background effects
in the Compton backscattering, particularly $e^+e^-$ pair
production in the collision of the laser with the high energy
photon in the conversion region,
$x_{1,min}=0$, $x_{2,max}=1$,
$x_{2,min}=\frac{\tau}{0.83}$,
$\hat{s}_{min}=m^2_{\tilde{e}}+m^2_d+m_{\tilde{e}} m_d$.
Then we can write the total cross-section as :

\begin{eqnarray}
\sigma=\int^{0.83}_{ {m^2_{\tilde{e}}+m^2_d+m_{\tilde e} m_d}/s }
d\tau\int^{1}_{\tau/0.83}
dx\frac{1}{x}f_{\gamma}(\frac{\tau}{x})f_q(x)\hat{\sigma}
(\tau s,m_{\tilde{e}})
\end{eqnarray}
where $f_q(x)$ is the distribution of up-quarks inside
the proton \cite{eich84}
\begin{eqnarray}
f_q(x)=2.751 x^{-0.412}(1-x)^{2.69}
\end{eqnarray}
and $f_{\gamma}(y)$ is the energy spectrum of the high energy
real photons (Ginzburg et.al. in ref. \cite{aydin96})
\begin{eqnarray}
f_{\gamma}(y)=\frac{1}{D(\kappa)}\biggr[1-y-\frac{1}{1-y}-\frac{4y}
{\kappa(1-y)}+\frac{4y^2}{{\kappa}^2(1-y)^2}\biggr]
\end{eqnarray}
with $y=\frac{E_{\gamma}}{E_e}, \kappa\cong 4.8, D(\kappa)
\cong 1.84$,and $y_{max}\cong 0.83$. $Q^2$ independent proton structure
function used above is satisfactory for the present analysis.

Since the third generation sfermions are usually expected to be lightest
we consider $\lambda'_{31k}$ couplings hence look for the production of
$\tilde {\tau}$. Taking
$\lambda'_{311}=0.11 \times \frac {m_{\tilde d}}{100 GeV}$ as
given in the literature the results of the numerical integration
for $\gamma p \rightarrow \tilde{\tau}_L d_R X$
is plotted in Fig.1.

Similarly the d-quark inside the proton permits the production
$\gamma p \rightarrow \tilde{\tau}_L u_R X$ through again the
$\lambda'_{311}$ coupling which
contributes roughly half of the u-quark contribution
and leads to the same  signature.
As can be seen from the figure the total cross-section for
a $\tilde{\tau}$ mass of 300 GeV is about 0.1 $pb$. Hence around 100
events per running year can be seen at HERA+LC up to
$\tilde{\tau}$ masses of  300 GeV. For comparison we note that
for the resonant production of
squarks of masses up to 200 GeV at  the HERA the total cross-section is
0.1 - 1 $pb$ (see E.Perez et.al. in ref.[10]).
On the other hand, if one uses the Weizsacker-Williams
approximations for the quasi-real photon distribution at the HERA machine
a similar process is possible but with
an almost hundred times smaller cross-section since WW-spectrum
is much softer than the real $\gamma$-spectrum. Clearly for the HERA machine
resonant production of the sparticles in R-parity violating
MSSM is the dominant process.\\
{\bf {Signature:}}
In models with R-parity violation the LSP is unstable, which leads
to signatures which differ strongly from the characteristic
missing energy signals in usual MSSM. In our case
the produced slepton ($\tilde{\tau}$) will decay either by direct
${R\!\!\!/}_p$ couplings as
$\tilde e^i_L \rightarrow e^j_R +\nu_k$  (through $ \lambda_{ijk}$)
or $d^j_R+u^k_L$ (through $\lambda'_{ijk}$) leading to the signals
1 $lepton$ + 1 $jet$ + ${E\!\!\!\!/}_T$ or 3 $jets$; or by cascading
through MSSM to the LSP which in turn decays via  ${R\!\!\!/}_p$:
$\tilde e^i_L \rightarrow e^i_L + \tilde{\chi}^o_1$,
$\tilde{\chi}^o_1 \rightarrow \ell^+\ell^-\nu
 \  {\rm or} \  q'\bar q\ell$
leading to the signals 3 $leptons$ + 1 $jet$ + ${E\!\!\!\!/}_T$ or
2 $leptons$ + 3 $jets$.
These LSP decays depend on couplings $\lambda '$ but also on the
supersymmetry parameters $M_2$, $\mu$ and $\tan {\beta}$
(see E.Perez et.al. ibid). In  the special cases involving only
the operators $L_iQ_j\bar D_j$ the LSP can also dominantly decay
via the radiative process
$\tilde {\chi}^o_1 \rightarrow \gamma + \nu $ \cite{dreiner91}.
Another possibility within the MSSM is
first decaying to a chargino by
$\tilde e^i_L \rightarrow \nu_i + \tilde{\chi}^\pm_1$ and then
followed by the chargino decay through
 $\tilde{\chi}^\pm_1  \rightarrow
 \ell^\pm q \bar q $ or $ \nu q {\bar q}'$ leading to the signals
1 $lepton$ + 3 $jets$ + ${E\!\!\!\!/}_T$ or
3 $jets$ + ${E\!\!\!\!/}_T$. One has also the decay
$\tilde{\chi}^\pm_1 \rightarrow W^\pm \tilde Z^o $ leading to
$multijets$ + $leptons $ + ${E\!\!\!\!/}_T$.
The ${R\!\!\!/}_p$ decays of $\tilde{\chi}^\pm_1$ dominate
over MSSM decays as long as $\lambda' $ is not too small
(e.g. around 0.1).
 On the other hand the above ${R\!\!\!/}_p$ decays of the produced
slepton have the following partial decay widths :
\begin{eqnarray}
\Gamma_{\tilde e^i_L \rightarrow e^j_R + \nu_k}=\frac
{m_{\tilde e}(\lambda_{131})^2}{16\pi}\\
\Gamma_{\tilde e^i_L \rightarrow d^j_R + u^k_L}=\frac
{m_{\tilde e}(\lambda'_{311})^2}{16\pi}
\end{eqnarray}
Therefore since $(\lambda'_{311})^2 \gg (\lambda_{131})^2$,
$\tilde {\tau}_L \rightarrow d_R+u_L$
is the main ${R\!\!\!/}_p$ decay leading to 3 $jets$.

\section{Single sneutrino production at gamma-proton colliders}
\label{sec:sneutr}
${R\!\!\!/}_p$ Yukawa couplings $\lambda'_{ijk}$ offer also the
opportunity to produce single sneutrino in gamma-proton collsions.
The relevant subprocess
$\gamma d \rightarrow \tilde{\nu}^i_L d^k_R$,
proceeds via the d-quark exchange
in s- and t-channels. The invariant amplitude is given as in Equ.(5),
with $Q^{\mu}$ as,

\begin{eqnarray}
Q^{\mu}={1\over2} (1-\gamma_5)\biggr [
\frac{\not\! k+\not\! p}{\hat s-m^2_d}+
\frac{\not\! p'-\not\! k}{\hat t-m^2_d}
\biggr ]\gamma ^{\mu}
\end{eqnarray}

Since in the t-channel instead of a heavy sparticle a d-quark is
exchanged the cross section for sneutrino production
becomes considerably bigger than the slepton case. The details
of the calculation for the total cross section is similar to the
one in the previous process. Taking
$\lambda'_{311}=0.11 \times \frac {m_{\tilde d}}{100 GeV}$ and
the simplistic $Q^2$ independent
distribution of down-quarks inside the proton as [11]
\begin{eqnarray}
f_q(x)=0.67 x^{-0.6}(1-x^{1.5})^{4.5}
\end{eqnarray}
the results of the numerical integration for
$\gamma p \rightarrow \tilde{\nu}_{\tau L} d_R X$ is plotted
in Fig.2. Hence around 100 events per running year can be seen at
HERA+LC up to $\tilde{\nu}_\tau$ masses of 175 GeV. For single
sneutrino production at LEP and hadron colliders see
\cite{lep99},\cite{barbier98} and \cite{moreau00}.\\
{\bf {Signature:}} The produced sneutrino can directly decay
to the ordinary particles through
$\tilde {\nu} \rightarrow \ell^+ \ell^-$ via $LL\bar E$ interactions
or to $\tilde {\nu} \rightarrow q \bar q'$ via $\lambda'_{ijk}$
couplings. Assuming dominant $\lambda'_{ijk}$ coupling constants then
the principal ${R\!\!\!/}_p$ decay signature is $three$ $jets$.
On the other hand sneutrinos may have gauge decays,
$\tilde {\nu}_i \rightarrow \chi^\pm \ell_i$ also. The produced
chargino can  decay into a neutralino which has further possible
${R\!\!\!/}_p$ decay modes noted in the previous section leading to
$three$ $jets$ + $three(two)$ $leptons$ + missing energy.

\section{Single sneutrino production at electron-photon colliders}

${R\!\!\!/}_p$ Yukawa couplings $\lambda_{[ij]k}$
(i.e. fourth term in Equ.(4) ) offer the opportunity
to produce sneutrinos in photon-electron collisions.
One of the the relevant subprocess
$\gamma e \rightarrow \tilde{\nu}_{\tau L} \mu_R$,
proceeds via electron (left-handed) exchange
in s-channel and muon (right-handed) exchange in t-channel.
 The invariant amplitude in two-component MSSM language is given as

\begin{eqnarray}
M=g_e \lambda \epsilon_{\mu}(k)\psi_+(p')Q^{\mu} \psi_-(p)\\[4mm]
Q^{\mu}=\biggr [\frac{(k+p)_\nu}
{\hat s-m^2_e}\sigma ^{\nu} \bar{\sigma} ^\mu+
\frac {(p'-k)_{\nu}}{\hat t -{m_{\mu}}^2}\sigma ^{\nu} \bar{\sigma}^\mu
\biggr ]
\end{eqnarray}
where $\psi_+(p')$ ($\psi_-(p)$) is the Weyl spinor for the right-handed
muon (left-handed electron) and $\lambda_{132}=0.06$
(present upper bound) for this particular
process. The total cross section for the above $\gamma e$ process may be
written as

\begin{eqnarray}
\hat{\sigma} (\hat{s},\gamma e)= \int^{t_{max}} _{t_{min}}
\frac{1}{16\pi\hat{s}^2}M^2 d\hat t
\end{eqnarray}
where
\begin{eqnarray}
t_{max/min}=(\frac{m^2_{\tilde {\nu}}-m^2_{\mu}}{2 \sqrt{s}})^2
-\biggr \{ \frac{\sqrt s}{2} \mp
\biggr [ \frac {(s+m^2_{\mu}-m^2_{\tilde \nu})^2}{4s}
- m^2_{\mu} \biggr ]^{1/2}
\biggr \}^2
\end{eqnarray}
The total cross-section for $e^+ e^- \rightarrow e \mu \tilde {\nu_\tau}$
is given by

\begin{eqnarray}
\sigma (s, e^+ e^-)=2 \int^{0.83}_{ {m^2_{\tilde{\tau}}}/s }
f_{\gamma /e}(y)\hat{\sigma}
(ys,m_{\tilde{\nu}})dy
\end{eqnarray}
where $f_{\gamma /e}$ is the distribution of high energy real photons
at a given fraction $y=\frac{E_{\gamma}}{E_e}$. The factor two is due to
the anti-sneutrino coming from the charge conjugate diagrams. The results
of numerical integration for $e^+ e^-$ $\rightarrow$ $e\mu \tilde{\nu}_\tau$
is plotted in Fig.3 for a linear collider with center-of-mass energy of 500 GeV.
As can be seen from the figure around $one$ $thousand$ events per running year
can be observed up to $\tilde {\nu}_\tau$ masses of 200 GeV. The same process
has been investigated in \cite{allanach97} for  LEP2 and NLC but with
Weizsacker-Williams approximation for the photon. They also present
detailed background analysis, however their numerical results concerning the
total cross section are
somewhat ambigous because the lower limit in Equ.(19) is taken to be zero
which affects the results severely.

In this paper we have first investigated  single slepton and
sneutrino productions at TeV scale $\gamma p$ colliders considering
R-parity violating $LQ\bar D$ interactions. Also sneutrino production
through $LL\bar E$ interactions at $\gamma e$ mode of the
NLC colliders has been discussed.
The production cross sections are
functions of only sparticle mass and ${R\!\!\!/}_p$
coupling constants, and
lead to detectable signals. Polarizations of initial photon
beams [8], which can be accomplished relatively easily,
constitute additional advantages. Our results
show that $\gamma e$ and $\gamma p$ colliders can play complementary role
in searching for supersymmetry in the future.

\acknowledgments

One of the authors, Z.Z.Aydin, thanks Alexander von Humboldt
Foundation for its support.

\newpage

\widetext

\smallskip

\figure{ Production cross section of
stau as a function of its mass for HERA+LC
$\gamma p$ collider.\label{fig1}}

\figure{ Production cross section of
sneutrino as a function of its mass for
HERA+LC $\gamma p$ collider.\label{fig2}}

\figure{ Production cross section of
sneutrino as a function of its mass for
NLC $\gamma e$ collider.\label{fig3}}

\begin{center}
\input{figstau}
\smallskip
\smallskip

Fig.1
\end{center}
\bigskip

\begin{center}
\input{figsngp}
\smallskip
\smallskip

Fig.2
\end{center}
\bigskip

\begin{center}
\input{figsnge}
\smallskip
\smallskip

Fig.3

\end{center}

\smallskip

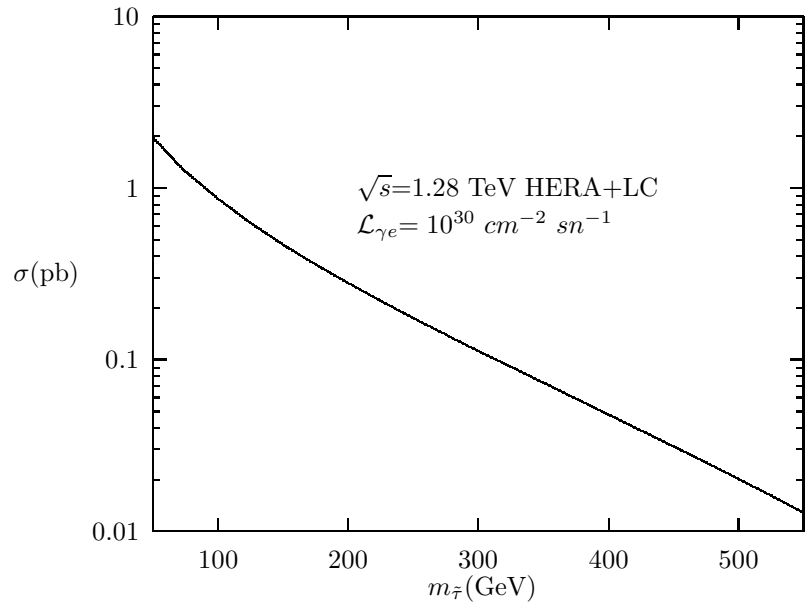
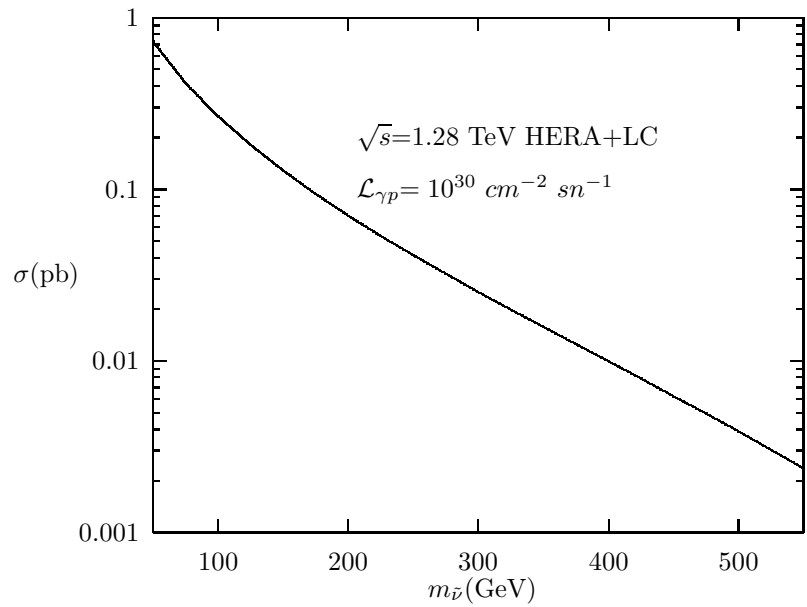
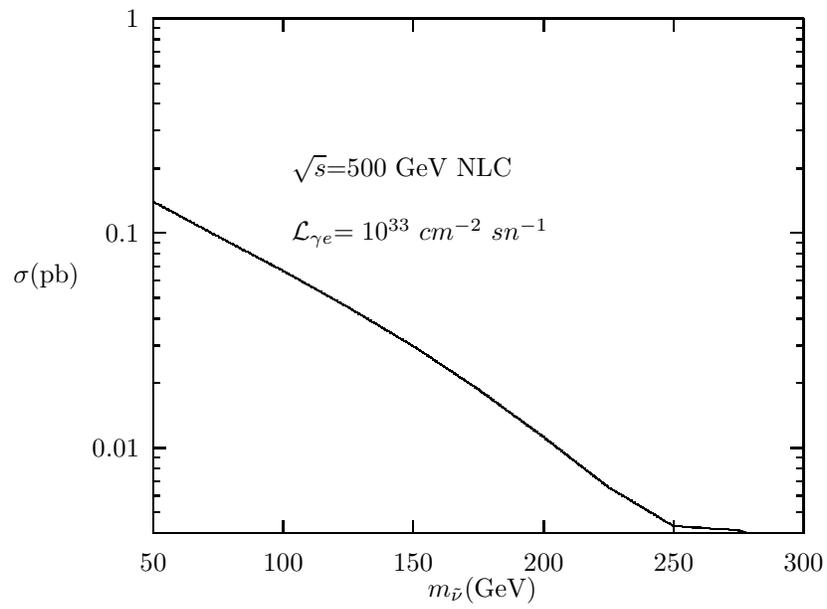
\end{document}